\documentclass[twocolumn,showpacs]{revtex4}
\usepackage[dvips]{graphicx}
\usepackage{amsmath,psfrag}
\usepackage{url} 

\usepackage[ps2pdf]{hyperref} 
\hypersetup{colorlinks=false, pdfpagemode=UseOutlines,
 bookmarksopen=false, bookmarksnumbered=true, pdfstartpage= {1},
 pdftitle    = {Boolean networks with scale-free in-degree distribution}
 pdfsubject  = {Work in process},
 pdfauthor   = {florian.greil<at>physik.tu-darmstadt.de},
 pdfkeywords = {Institut fuer Festkoerperphysik, Technische Universitaet Darmstadt, Germany},
 pdfcreator  = {Adobe-Acrobat-Distiller},
 pdfproducer = {LaTeX with hyperref}
}

\begin{document}
\title{Critical Boolean networks with scale-free in-degree distribution}
\author{Barbara Drossel, Florian Greil}
\email[]{\tt florian.greil@physik.tu-darmstadt.de}
\affiliation{Institut f\"ur
Festk\"orperphysik, Technische Universit\"at Darmstadt, 64289 Darmstadt, Germany}
\email[]{\tt http://www.fkp.tu-darmstadt.de/drossel}
\begin{abstract}
  We investigate analytically and numerically the dynamical properties
  of critical Boolean networks with power-law in-degree distributions.
  When the exponent of the in-degree distribution is larger than 3, we
  obtain results equivalent to those obtained for networks with fixed
  in-degree, e.g., the number of the non-frozen nodes scales as
  $N^{2/3}$ with the system size $N$.  When the exponent of the
  distribution is between 2 and 3, the number of the non-frozen nodes
  increases as $N^x$, with $x$ being between 0 and 2/3 and depending
  on the exponent and on the cutoff of the in-degree distribution.
  These and ensuing results explain various findings obtained earlier
  by computer simulations.
\end{abstract}
\maketitle

Complex dynamical systems, where a large number of units interact in a
non-trivial way, are often modelled as networks. The units from
which these networks are built can show various types of intrinsic
dynamics, including oscillations. Whenever the dynamics can be reduced
to only two possible states per node, a Boolean network is obtained.
Stuart Kauffman was the first to use Random
Boolean networks (RBNs) to model the dynamics of genetic and protein
networks~\cite{kauffman:metabolic,kauffman:homeostasis}. Although
Boolean models represent a strong simplification of the far more complex
reality, there exist several examples where the modelling of a cellular
network by Boolean variables captures correctly the essential dynamics
of the system \cite{bornholdt:less,albert:topology}.

RBNs are directed graphs where each node~$i$ has a Boolean
value~$\sigma_i \in \{0,1\}$ and an update function~$f_i$ which
determines the new value in the next time step as function of the
state of those nodes that have a link to node $i$. Links and 
functions are assigned at random, given certain constraints concerning
the number of inputs per node or the set of functions.  The update can
be performed in different ways, we consider here the usual case of
synchronous update. After some time, the dynamics reaches an
attractor, i.e. a periodic sequence of states. Depending on the
parameters of the network, the dynamics is either in the \emph{frozen}
phase or in the \emph{chaotic} phase or at the critical point between
the two. In the frozen phase, all apart from a small number of nodes
assume a constant value on the attractors, i.e., they are frozen. When
the state of a node is changed, on average less than one node will be
changed in the next time step, and the size of a perturbation
decreases with time. In the chaotic phase, a nonvanishing proportion
of nodes keep changing their state even after a long time.  The size
of a perturbation increases with time, since a change in the state of
one node will lead on an average to a change of the state of more than
one nodes in the next time step.  Most studies of RBNs have focused on
the critical point, which is at the boundary between these two phases,
and where a perturbation of one node propagates on an average to one
other node.  These studies deal mainly with the (mean) number and size
of attractors, motivated by Kauffman's original claim that biological
networks are poised at the critical point, and that attractors can be
equated with cell types. Despite of the long time since the
introduction of the model, a full analytical understanding of critical
RBNs was obtained only recently
\cite{samuelsson:superpolynomial,kaufman:scaling,drossel:review}.

A key concept at understanding the dynamics of critical RBNs is the
classification of the nodes according to their dynamical behavior on
attractors into frozen, non-frozen and relevant nodes
\cite{bastolla:modular}. Relevant nodes are those nodes that determine
the attractors, while the other nonfrozen nodes are slaved to the
dynamics of the relevant nodes; changing their state does not change
the attractor.  A stochastic process that gradually determines the
frozen core starting from the nodes that have a constant function, was
used in \cite{kaufman:scaling,mihaljev:scaling} to prove that the
number of nonfrozen nodes in critical RBNs scales as $N^{2/3}$, and
the number of relevant nodes as $N^{1/3}$, with $N$ being the number
of nodes in the network.  In the limit of large network size, scaling
functions for the number of non-frozen and relevant nodes were
calculated analytically.  

All studies mentioned so far assign $k$ inputs to each node, while the
number of outputs is Poisson distributed, since incoming links are
connected at random to a node where they originate. However,
biological networks are known to have a broad degree distribution,
which is often well described by a power law (see
\cite{albert:scale-free} and references therein).  For this reason,
several recent studies were devoted to Boolean dynamics on scale-free
networks. The majority of these studies uses a scale-free in-degree
distribution and a Poissonian out-degree distribution, but other
implementations can also be found. Observations made in computer simulations are
that attractors are shorter and frozen nodes are more numerous in
critical scale-free networks compared to RBNs with a fixed number of
inputs, given the same total number of links and of nodes
\cite{fox.hill:from,kinoshita.iguchi.ea:robustness}, and that
attractors are sensitive to perturbations of highly connected nodes,
but not of sparsely connected nodes
\cite{aldana:boolean,kinoshita.iguchi.ea:robustness}. These and other
\cite{silva.silva.ea:scale-free,serra.villani.ea:on} simulation
results are merely stated and are not embedded into an analytical
framework.  Analytical results obtained so far are limited to
calculating the phase diagram using the annealed approximation
\cite{aldana:boolean,aldana.cluzel:natural,fronczak.fronczak.ea:kauffman};
only the work by Lee and Rieger \cite{lee.rieger:broad} goes further
by calculating the asymptotic Hamming distance in the chaotic phase
and extrapolating the results to the critical point by using a
finite-size scaling ansatz in combination with the calculation of the
size distribution of perturbed clusters.

In this paper, we will present an analytical calculation for RBNs with
scale-free input distributions at the critical point, obtaining
scaling laws for the number of nonfrozen and relevant nodes. Our
results, which are confirmed by a numerical evaluation, explain the
above-mentioned findings of computer simulations, and convey a clear
understanding of the properties of attractors in these systems.

We consider critical networks that have an in-degree distribution
$P(k)$ that follows a power law, $P(k) = A \cdot k^{-\gamma}$ for
$k>1$.  The normalization constant~$A$ depends on the minimum and the
maximum in-degree. We fixed the minimum in-degree to 2; the
maximum in-degree depends on the network size and the chosen
implementation of the model (see below). We consider only the case
$\gamma > 2$, where such a normalization is possible.  In the case $2
< \gamma < 3$, the second moment of the degree distribution diverges,
and it has been argued in \cite{lee.rieger:broad} that this should
change the dynamical properties.  The out-degree distribution is
Poissonian with a mean $\langle k \rangle=\sum_k kP(k)$, since the
input connections are chosen at random from all nodes, just as for
RBNs with fixed $k$.

We investigated two ways of creating the input distribution.  First,
we assigned to each node $i$ a number $k_i$ of inputs that was drawn
from the distribution $P(k)$, not allowing values $k_i$ larger than
$N$ or smaller than 2. The total number of links and the largest value
of $k_i$ differ in this case between different networks. Second, we
fixed the number of nodes with $k$ inputs exactly at the value $NP(k)$
(rounded to the nearest integer), which gives a distribution $P(k)$
that has a cutoff at $k \sim N^{1/\gamma}$. In part of the
above-mentioned studies, networks with scale-free in-degree
distributions were generated using a constraint that does not allow
multiple connections between the same nodes, or using a
preferential-attachment algorithm, however, all these are known to
create correlations between the degree of neighboring nodes
\cite{weber.porto:generation}, which in turn can affect the dynamics on
these networks \cite{weber.huett.ea:pattern}. In order to avoid such
complications, we connect the incoming links at random to any node,
without imposing any constraints. 

We also investigated several ways of assigning the Boolean functions
to the nodes. First, we chose biased functions with a parameter $p$,
assigning to each of the $2^{k_i}$ input configurations the output 1
with a probability $p$ and the output 0 with a probability $1-p$. The
value of $p$ was chosen such that the network is critical, i.e., that
$p=1/\langle k \rangle$ \cite{aldana:boolean}. The main results
did not depend on whether we chose the exact mean (which can be
different for each network), or the theoretical mean $\sum_k k P(k)$.
The second way of assigning the Boolean functions is to take only
constant and reversible functions.  There are two constant functions,
which fix the value of a node to either 0 or 1, irrespective of its
input values.  For each value of $k$, there are 2 reversible
functions, which are defined by the condition that changing the value
of one input always changes the output. A node with a reversible function
becomes frozen only if all of its inputs are frozen. Such a network is
critical if the total number of nodes equals the total number of
inputs to nodes with reversible functions. Links to nodes with
constant functions have no effect and can be omitted, so that the
total number of links becomes identical to the total number of nodes.

\begin{figure}[ht!]
\begin{center}
\psfrag{C0}{$C_0$} \psfrag{C1}{$C_1$} 
\psfrag{C2}{$C_2$} \psfrag{C3}{$C_3$}
\includegraphics[width=0.44\textwidth]{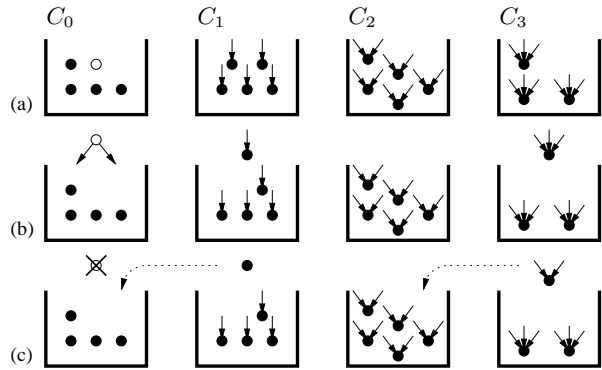}
\end{center}
\caption{A sketch of one step of the stochastic process to determine
the frozen core. (a) Nodes are placed in containers according to the
number of inputs of which we do not yet already know for sure that
they are frozen. We choose a node~$\circ$ from the container $C_0$.
(b) This node becomes an input to a node in container $k$ with
probability $k/N$. In this example, it becomes the input to 2 nodes.
(c) The frozen links are removed, the two nodes are moved to
container $C_{k-1}$, and node~$\circ$ is removed from the system.}
\label{ProcessSketch}
\end{figure}

We adjusted the method proposed in \cite{kaufman:scaling} in
order to determine the size of the frozen core of critical networks
with scale-free input distributions. In the following, we describe this
method for the case of only constant and reversible functions. The
generalization to other cases is straightforward.
The frozen core is determined starting from the nodes with constant
functions and determining stepwise all those nodes that become frozen
because all their inputs are frozen.  The main idea of our method is
to not specify the network in advance, but to choose the connections
within the network while determining the frozen core. To this purpose,
we place all $N$~nodes of the network into ``containers'' $C_k$
according to the number $k$ of inputs.  As mentioned above, inputs to
nodes with constant functions are omitted, and these nodes are
therefore put into container $C_0$. The largest container index is
$k_{\rm max}=N$ or $k_{\rm max} \sim N^{1/\gamma}$, depending on the
method chosen for creating the input distribution.  The contents of
the containers change with time, since we remove stepwise all those
inputs of which we know that they come from a frozen node.  The
``time'' we are defining here is not the real time for the dynamics of
the system, but it counts the steps of the stochastic process that we
use to determine the frozen core. During one time step, we choose one
node from the container $C_0$ and determine to which nodes this node
is an input.  Since the inputs are picked at random, the chosen node
is connected to each input with probability $1/N$.  These inputs are
removed, and the corresponding nodes moved from container $C_k$ to
container $C_{k-1}$ (or to a lower container, when more than one
connection is made to the same node).  At the end of the time step, we
remove the chosen node from the system, and the number $N$ of nodes in
the system is reduced by 1.  Thus, at each time $t$ the number $
|C_k|$ of nodes in container $C_k$ is the number of nodes that have
$k$ inputs that have not yet become frozen during the process.  In
container $C_0$ are those nodes of which we know already that they are
frozen, but we have not yet determined to which other nodes they are
an input. We denote from now on the total number of nodes in the
system by $N_{\rm ini}$, which is identical to $N(t=0)$. The number
$N_{\rm ini}-N(t)$ of nodes have been removed from the system. They
are those nodes for which we have already determined that they are
frozen and to which other nodes they are an input.

The process ends when there are no nodes left in container $C_0$, or
when all nodes are in container $C_0$. In the latter case, the entire
network freezes, and the dynamics of the system runs to the same fixed
point for all initial conditions. In the first case, there is a set of
nonfrozen nodes. In order to determine the topology of the nonfrozen
part of the network, one can then fix the connections that have not
yet been determined by connecting the remaining inputs at random to
the remaining nodes.

Before showing the results of our computer simulations of this process
obtained for an ensemble of many networks, let us first perform an analytical
calculation in order to predict the mean number of nodes remaining in
the different containers at the end.  We begin by evaluating the mean
number of nodes in container $C_k$ at the moment where only the
fraction $\epsilon=N/N_{\rm ini}$ nodes are left in the system. At
this moment, container $C_k$ contains all nodes that had initially
$l\ge k$ inputs, and where $l-k$ inputs have already become frozen.
The probability that an input has not yet become frozen is identical
to $\epsilon$, since only the proportion $\epsilon$ of nodes have not
yet been removed, and since an input is connected to every node with
the same probability. Since container $C_l$ contained initially
$\propto N_{\rm ini} l^{-\gamma}$ nodes, we have
\begin{eqnarray}
|C_k| &\propto& N_{\rm ini} \sum_{l=k}^{k_{\rm max}} 
l^{-\gamma} \! \epsilon^k \! (1- \epsilon)^{l-k} \! {l \choose k}\, .
\end{eqnarray}
For small $\epsilon$, nodes in container $C_k$ originated in
containers $C_l$ with $l \gg k$, and we can therefore set $l-k \approx
l$. Replacing the sum with an integral, using $e^{-x} \simeq
(1-x)$ and ${l \choose k} \simeq l^k$, we obtain the approximate
expression
\begin{eqnarray}
|C_k|&\propto & N_{\rm ini} \epsilon^k \!
\int_k^{k_{\rm max}} \!\!\!\! l^{k-\gamma}
e^{- l\epsilon} {\rm d}l \label{main}
\end{eqnarray}

When evaluating this integral, we have to consider three possible cases:
\begin{enumerate}
\item The integral is independent of the cutoff because $k<\gamma -1$. In this
case we obtain
\begin{equation}
|C_k| \sim  N_{\rm ini} \epsilon^k \, . \label{Case1}
\end{equation}
\item  $k>\gamma -1$ and  $\epsilon^{-1}<k_{\rm max}$: in this case
the exponential function determines the cutoff to the integral, and we
obtain
\begin{equation}
|C_k| \sim  N_{\rm ini} \epsilon^{\gamma -1} \, . \label{Case2}
\end{equation}
\item  $k>\gamma -1$ and  $\epsilon^{-1}>k_{\rm max}$:  in this case
$k_{\rm max}$ determines the cutoff to the integral, and we obtain
\begin{equation}
|C_k| \sim  N_{\rm ini} \epsilon^k  k_{\rm max}^{k-\gamma+1} \, . \label{Case3}
\end{equation}
\end{enumerate}

The stochastic process ends when no nodes are left in container $C_0$.
On an average, the number of nodes in container $C_0$ is identical to
the number of nonfrozen inputs minus the number of nonfrozen nodes,
since the network is critical. If we neglect stochastic fluctuations
during the process, the number of nodes in container $C_0$ becomes
zero at the same time when the number of nodes in container $C_k$ with
$k>1$ becomes zero, i.e. when $\epsilon=0$. However, stochastic
fluctuations will terminate the process earlier, at the moment where
the fluctuations of the number of frozen nodes become of the same
order as the expected number of frozen nodes. The variance of the
number of frozen nodes is evaluated as follows: The probability that a
given input has not yet become frozen at the moment where $N$ nodes
are left in the system, is $\epsilon$. When $\epsilon$ is small, the
number of nonfrozen inputs is Poisson distributed, with the variance
being identical to the mean, which is proportional to $N_{\rm
  ini}\epsilon$.  For small $\epsilon$, the vast majority of nonfrozen
inputs is found in container $C_1$. Now a node in container $C_1$
would be in container $C_0$ had its remaining input also become frozen
during the process, and it follows that the variance of the number of
frozen nodes is also of the order $N_{\rm ini}\epsilon$.  The typical
fluctuations in the number of frozen nodes are therefore of the order
$\sqrt{N_{\rm ini}\epsilon}=\sqrt{N}$.  Equating this number with the
expected number of nodes in $C_0$, which in turn is of the same order
as the expected number of nodes in $C_2$, we obtain the following
condition for the end of the stochastic process, where $N$ is
identical to the number of nonfrozen nodes, $N_{\rm nf}$:
\begin{equation}
|C_2| \sim  \sqrt{N_{nf}} = \sqrt{N_{\rm ini}\epsilon} \, . \label{end}
\end{equation}

\begin{figure*}[bt!]
\begin{center}
\includegraphics[width=0.9\textwidth]{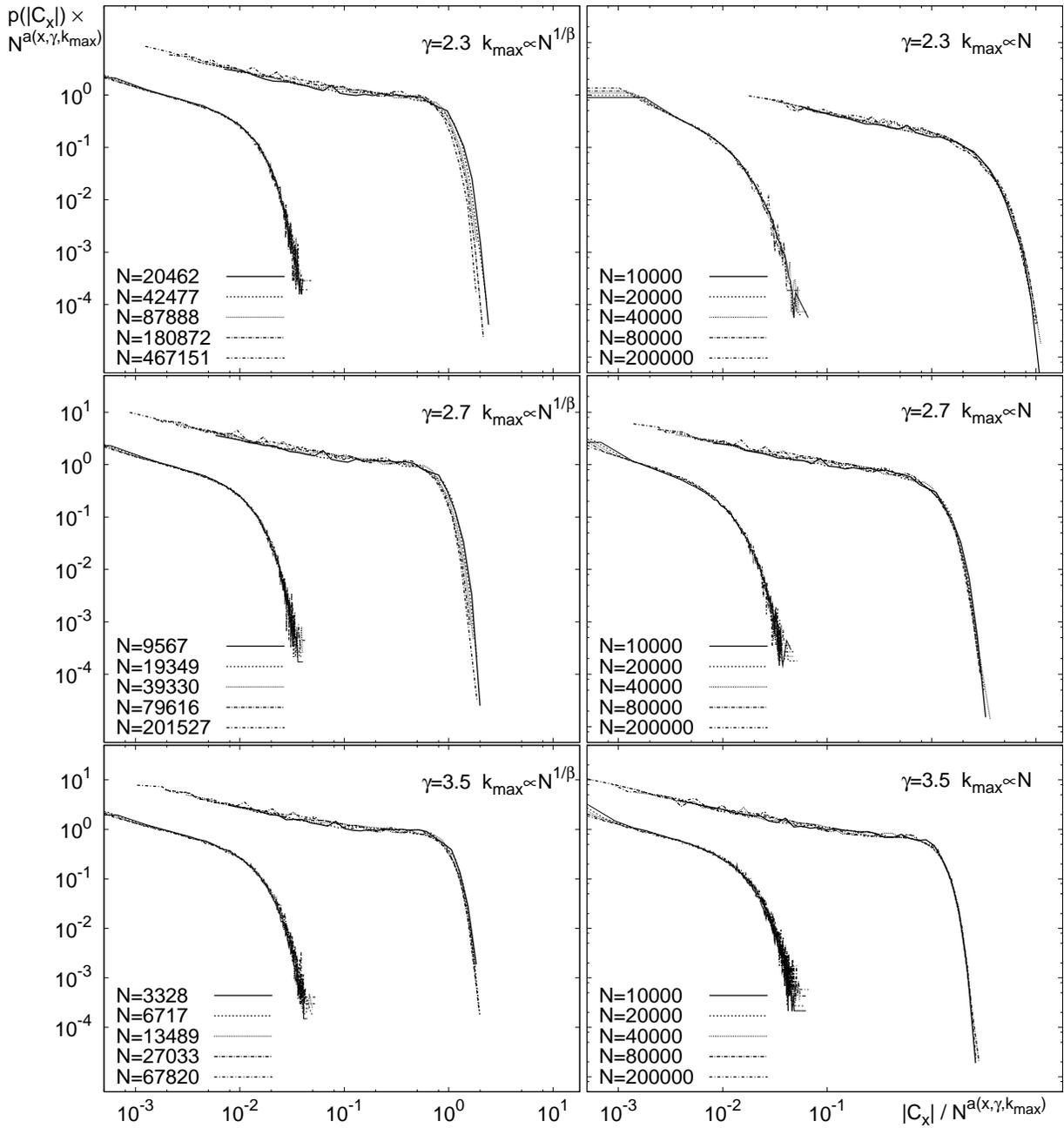}
\end{center}
\caption{Scaling collapse for the total number of nonfrozen nodes
  ($x=1$) and for the number of nodes with two nonfrozen inputs
  ($x=2$, left curves in each graph, shifted to the left by a factor
  10 for better visibility), for three different values of $\gamma$
  and for the two different ways of choosing the input
  distributions. The function $a(1,\gamma,k_{\rm max})$ is the
  appropriate exponent in Eqs.~(7) to (9), and $a(2,\gamma,k_{\rm
  max}) = a(1,\gamma,k_{\rm max})/2$. Each data set is generated by
  averaging over $10^5$~realizations. }
\label{FigSim}
\end{figure*}

Depending on the value of $\gamma$ and on the dependence of $k_{\rm max}$ on
$N_{\rm ini}$, the number of nonfrozen nodes scales in a different way
with $N_{\rm ini}$.  

For $\gamma > 3$, the first of the three above cases applies to $|C_2| $, and
solving condition (\ref{end}) for $N_{\rm nf}$ we obtain 
\begin{equation}
N_{\rm nf} \sim N_{\rm ini}^{2/3}
\end{equation}
at the end of the stochastic process.  This is the same result as for
a RBN with fixed $k$. Whenever the input distribution $P(k)$ has a
finite second moment, the number of nonfrozen nodes scales as
$N_{\rm ini}^{2/3}$, and the number of nonfrozen nodes with two nonfrozen inputs
scales as $N_{\rm ini}^{1/3}$. The number of nonfrozen nodes with more than two
nonfrozen inputs depends on whether $k<\gamma-1$, but it is in any case
much smaller than the the number of nonfrozen nodes with two nonfrozen
inputs, and we do not evaluate it here further.

When $2<\gamma <3$ and when $k_{\rm max} \propto N_{\rm ini}$, the second case applies, and
we obtain using Eq.~(\ref{Case2})
\begin{equation}
N_{\rm nf} \sim N_{\rm ini}^{(2\gamma-4)/(2\gamma-3)}\, .
\end{equation}
For $\gamma = 3$, the exponent is $2/3$, $N_{\rm  nf}\sim N_{\rm ini}^{2/3}$, and it
decreases to 0 as $\gamma$ approaches 2. 

When $2<\gamma < 3$ and when $k_{\rm max} \propto N_{\rm ini}^{1/\gamma}$ (which is
the case when the input distribution is fixed), the third case
applies, and we obtain
\begin{equation}
N_{\rm nf} \sim N_{\rm ini}^{2\gamma/(\gamma+6)}\, . \label{eq9}
\end{equation}

These results Eqs. (\ref{end}) - (\ref{eq9}) should be also valid when biased Boolean functions are
chosen. In this case, there is a nonvanishing probability that a node
in container $C_k$ with $k>1$ becomes frozen when an input becomes
frozen.  Therefore the expression (\ref{main}) for $|C_k|$ obtains an
additional factor $1-p^{2^k}-(1-p)^{2^k}$, which is never close to 0
and therefore does not change the scaling behavior of the integral.

Our computer simulations confirm all these analytical considerations.
As an example, we show in Fig.\ref{FigSim} results obtained for the
case of only constant and reversible functions, for both ways of
choosing the input distributions.  The excellent quality of the data
collapses confirms our analytical calculations.

Our results have a variety of implications. First, they show that many
properties obtained for critical networks with a fixed number of
inputs apply also to networks with a scale-free in-degree distribution
once the frozen nodes have been removed. In particular, the number of
nonfrozen nodes with more than one nonfrozen input scales as the
square root of the number of nonfrozen nodes. Only the dependence of
the number of nonfrozen nodes on the total number of nodes is changed
when $\gamma \in (2,3)$. We can therefore take over the results
obtained in \cite{kaufman:scaling} based on these properties of the
nonfrozen nodes. It follows in particular that the number of relevant
nodes in networks with a scale-free input distribution scales as the
square root of the number of nonfrozen nodes, and that the number of
relevant components is of the order of $\log{N_{\rm ini}}$, with all
but a limited number of relevant components being simple loops.  It
therefore follows again that the mean number and length of attractors
diverges faster than any power law with the network size. This
explains the finding in \cite{aldana:boolean} that the state-space
structure of critical RBNs with fixed $k$ and with a power-law input
distribution is similar.  Second, the number of nonfrozen nodes
decreases with decreasing $\gamma \in (2,3)$, because the exponent
becomes smaller. This explains why several authors have seen more
frozen nodes and shorter attractors in scale-free networks compared to
standard RBNs. Third, the set of nonfrozen and relevant nodes is
dominated by nodes with many inputs. This is due to the fact that each
input has the same probability of surviving the stochastic process
until the end. The average number of inputs of a node that has a
surviving link is proportional to $\int k^2 N(k) dk$, which is
dominated by $k_{\rm max}$ for $\gamma \in (2,3)$. When a relevant
node is perturbed, the attractor is changed with a large probability,
however when a frozen node is changed, the attractor changes with a
probability that vanishes in the limit $N \to \infty$. This explains
the findings in \cite{aldana:boolean,kinoshita.iguchi.ea:robustness},
that attractors respond sensitively mainly to perturbations of highly
connected nodes. Fourth, our results disagree with the finite-size
arguments in \cite{lee.rieger:broad}, which predict that the number of
nonfrozen nodes scales as $N_{\rm ini}^{(\gamma-1)/\gamma}$. This is
in our view due to the fact that an infinite (sustained) perturbation
has properties that are fundamentally different from those of finite
perturbations, in which case arguments based on finite-size scaling do
not work.

\paragraph*{Acknowledgment.} 
This work was supported by the German Research Foundation (Deutsche
Forschungsgemeinschaft, DFG) under Contract No. Dr200/4.

\end{document}